\documentstyle[11pt,newpasp,twoside,epsf]{article}
\def\edcomment#1{\iffalse\marginpar{\raggedright\sl#1\/}\else\relax\fi}
\reversemarginpar

\begin{document}
\title{Case Studies of Mass Transfer and Star Formation in Galaxy
Collisions} 
 \author{Curtis Struck}
\affil{Dept. of Physics \& Astronomy, Iowa State Univ., Ames, IA
50011  USA}
%\author{Co-authors}
%\affil{Their affiliation}

\begin{abstract}

The amount, timing and ultimate location of mass transfer and induced
star formation in galaxy collisions are sensitive functions of orbital
and galaxy structural parameters. I discuss the role of detailed case
studies and describe the results for two systems, Arp 284 and NGC
2207/IC 2163, that have been studied with both multiwaveband
observations, and detailed dynamical models. The models yield the mass
transfer and compressional histories of the encounters and the
``probable causes'' or triggers of individual star-forming
regions.

\end{abstract}

\section{Introduction: The Case for Case Studies}

There are two general approaches to the study of the complex effects
of galaxy collisions. The first is statistical, such as the study of
particular properties of a reasonable sample of interacting galaxy
systems, or a grid of numerical models covering some range of initial
conditions. The second approach is the case study, the detailed
investigation of a particular collisional system.

The first approach is the path more frequently followed, in the
literature. A great deal can be learned from observations of the
global properties of many systems, which are easier to acquire
(individually), than the high sensitivity, high resolution,
multi-wavelength data needed for a good case study. In the realm of
simulations, the statistical approach also has many attractions.  For
example, in many regions of parameter space the model outcomes are
quite sensitive to collision parameters, so converging on the correct
parameters can be a prolonged task. (At the same time, such
sensitivities can go a long way toward guaranteeing the uniqueness of
a successful model.)

One of the greatest successes of the statistical approach
(spurred by IRAS results) in the last two decades is an
understanding of how gas is redistributed in major mergers, and how
ultra-luminous, super-starbursts result.  Insights obtained from
numerical models (e.g., Barnes and Hernquist 1972) also played a
crucial role.

Statistical studies also provide important inputs to the topics of
mass transfer and induced star formation (SF) in galaxy
collisions. Even aside from studies of luminous major merger remnants,
there has been much work on the questions of whether and by how much
SF is enhanced by interactions. The early color analysis of Larson and
Tinsley (1978), suggested color dispersion enhancement, rather than
bluer colors. The nuclear spectrophotometry by Keel et al. (1985, also
Kennicutt et al. 1987) suggested enhancement of nuclear SFRs.

More recently, Bergvall et al. (2003 and earlier work cited therein)
find no difference in the broadband colors of sample of 59 interacting
systems relative to a control sample of 38 galaxies. They do find a
moderate increase in central SF and far-infrared emission in the
interacting sample. They argue that earlier work, claiming greater
enhancements was biased towards IR luminous merger remnants, in
contrast to their sample. Barton, Geller \& Kenyon (2003a, and Barton
Gillespie, Geller \& Kenyon 2003b) obtained B and R band photometry
and optical spectroscopy of 190 galaxies in pairs and compact
groups. They also found enhanced core SF, and a number of
post-starburst, as well as starburst cores. In addition they found an
anti-correlation between galaxy separation and SF, and suggested
starbursts were preferentially triggered at closest approach, decaying
thereafter.

These statistical studies shed the most light on nuclear SF,
suggesting that it has a short duty cycle, and perhaps requires a
relatively strong disturbance to funnel the gas fuel inward (Keel
1993). In this symposium we have seen some beautiful observations of
SF in tidal structures. The statistical studies suggest that such SF
does not add up to a very large amount. This echos the result of the
statistical study of Schombert, Wallin \& Struck-Marcell (1990) on the
colors of tidal bridges and tails. Yet, in tails it is the nature,
rather than the quantity of SF that is of interest. Case studies with
more details on the SF history, as well as instantaneous rates in
individual SF regions are needed to address the many unanswered
questions.

The processes of tidal or splash mass transfer are very hard to study
either individually, or statistically. We can observe the amount of
gas between or outside the interacting galaxies, but how much has been
transferred already, and how much will yet be pulled out? The
statistical question can be addressed with models, but there have been
few such studies. Wallin \& Stuart (1992) gave us a survey and
analysis of 1000 restricted 3-body encounters with mass transfer from
a particle disk around the primary. These models evenly sampled a
number of collision parameters, and quantified dependences, such as
inclination, some of which had been known since Toomre and Toomre 1972
(see Struck 1999, Sec. 4.1). Howard et al. (1993) published a nice
atlas of 86 N-body simulations (with rigid halos) of the effects of
encounters on (2-d) disks. I produced a small model grid to study the
hydrodynamics of colliding gas disks (Struck 1997).

Statistical studies show us the big picture, and give the average
answer to big picture questions, but often leave us wondering about
the specific mechanisms. Case studies provide detailed answers to some
of those questions, but they require a large amount of high quality
observational data, and a substantial modeling effort to interpret
it. As yet, not many have been published. However, we can now
optically resolve individual star clusters in nearby interacting
systems, map tenuous gas distributions, and we have the computer power
to do the modeling, so it is a great time for case studies. I will
justify that statement with two examples.

\section{Case 1: Arp 284 (NGC 7714/15)}

Bev Smith and I have been working for some years on this beautiful
system (see Figure 1, and Smith \& Wallin 1992, Smith et al. 1997,
Struck \& Smith 2003). 

\begin{figure}[h]
%\plotone{f1.ps}
%\plotfiddle{f1.ps}{1.5in}{0}{35}{35}{-120}{-80}
\caption{Arp atlas image of NGC 7714/15 (Arp 284, NGC 7714 is on the
right) from Struck \& Smith (2003). Feature 3 is the inner SW tail
discussed in the text.}
\end{figure}

\noindent It is an asymmetric collisional ring galaxy, with a
substantial bridge connecting it to a nearly edge-on companion, and it
also contains some distinctive tidal tails. It has a variety of
features that make it an especially interesting 'case' for the study
of induced SF and mass transfer. The primary has a prototypical
starburst nucleus, but their is little evidence for recent SF in the
ring wave, in contrast to most gas-rich collisional ring systems (see
Appleton and Struck-Marcell 1996). The companion has a post-starburst
spectrum. There are intriguing knots of recent SF in one of the
countertails, and roughly in a line on the north side of the
bridge. More than half of the gas mass is located outside the main
disks, in the bridge and tails (Smith et al. 1997). In the bridge,
there is an offset between the center line of the old stars, and that
of the gas.

Lan\c{c}on et al. (2001) have published detailed spectral evolutionary
models of HST observations, so we have a good picture of the SF
history of the nucleus of NGC 7714. In a word, this history seems to
be characterized by repetitive bursts.  Unfortunately, high resolution
HST spectra are not available for extra-nuclear regions, or for the
companion. (A. Lan\c{c}on is working to obtain VLT data, however.)

We have recently published N-body hydrodynamic (SPH) models of this
system that can account for nearly all of the observed morphological
and kinematic features (Smith \& Struck 2003). Successful models
require a high inclination collision, which is somewhat prograde for
galaxies. Such an orbit allows the simultaneous production of a
complete ring, and the tails.

In terms of mass transfer we found that we could indeed fling a large
fraction of the gas out of the two disks in such model collisions, and
without unusual initial conditions, like especially extended gas
disks. Specifically, our models put a somewhat large gas fraction than
observed into the great HI loop to the north, somewhat less than
observed in the bridge, and a good deal in the companion
countertail. This last feature is not isolated in the observations,
except as a slight eastern extension of the companion disk. The models
suggest that it lies behind both the companion disk and the bridge,
and cannot be easily distinguished.

Interestingly, we found that the bridge consists of several
superposed, but dynamically distinct components. In addition to the
two usual tidal components stretching from the near side of each
galaxy towards its companion, there is also the superposed countertail
of NGC 7715, and the remnants of an old tail winding around from the
far-side of NGC 7715 to the primary (NGC 7714). Mass transfer onto the
primary from this old bridge started before closest approach. Some of
the early transfer material was incorporated into the primary disk, or
in a plane very close to it, and subsequently flung out (again!) as
the inner SW countertail. This feature, and the fact that the location
and mass of some of the bridge components are very sensitive to
initial collision parameters, illustrate the intricacies of mass
transfer that can be found in detailed case studies.

Our models include simple feedback prescriptions that can tell us
something about induced SF. The gas compression history can also 
suggest SF phenomena even when we don't really have enough particles
to fully resolve them. Examples of the latter include the result that
the (star-forming) inner SW countertail of NGC 7714 may consist of
mass transferred material that has been compressed in the primary
disk. Similarly, the models show that the old bridge component is
overrun and shocked by a newer bridge component. This might account
for the line of young star clusters on the north side. The feedback
model further suggests that the companion experienced a starburst at
closest approach, consistent with its post-starburst spectrum. It also
suggests multiple bursts have occurred in the core of the primary, in
agreement with the spectral synthesis results.

However, the exact timing and number of these bursts are not
completely predictable. The models suggest that they could be driven
by compressions from the (m=0) ring wave component of the disturbance, and
fed by inward mass transfer resulting from the spiral component. On
the other hand, repetitive starbursts in the gas-rich core are driven
by the feedback terms even without an external disturbance. (This
result derives from control runs and other unpublished modeling.) It
is likely that the gas density in the core or its susceptibility to SF
was less than in the models.

\section{Case 2: NGC2207/IC2163}

I have been working on this galaxy with the 'ocular' galaxy
collaboration (see Elmegreen et al. 1991, 1995a,b, 2000, 2001, Kaufman
et al. 1997). The Hubble Heritage image of this graceful pair has been
often reproduced. The models suggest that it has been one of the
gentlest close interactions known. They further suggest that the
companion approached from slightly above the primary (in the west),
interacting with its outer edge, and then moving slightly below and to
the north and then east of the primary. The encounter is retrograde
relative to the primary, so its optical disk is not greatly perturbed,
though its larger HI disk is.

\begin{figure}[h]
%\plotone{f2.ps}
%\plotfiddle{f2.ps}{1.6in}{0}{30}{30}{-90}{-55}
\caption{NGC 2207/IC 2163 from Elmegreen et al. (2001, NGC 2207 is on
the right).}
\end{figure}

The encounter is prograde and nearly in-plane relative to the
companion, which is estimated to be about 60-80\% of the mass of the
primary. Encounters that strongly torque the disk produce long tidal
tails, and also the transient ocular (eye-shaped) morphology from
material that loses angular momentum (Elmegreen et al. 1991). This
system is a prototypical example of that process of disk
rearrangement. In addition, the models suggest that there is some
moderate mass transfer from the companion to the primary, both
presently and earlier in a glancing interaction on the west side. They
also suggest that the collisional perturbation did not produce the
spirals in the primary; they were almost certainly pre-existing.

Observationally, the recent star formation in this system is mostly
extended (and beautiful!) with no excess in the galaxy
cores. Specifically, the young star clusters are concentrated in the
long spiral arms. In the companion, the young clusters are
concentrated on the rim of the ocular. The tidal tail is younger and
shorter than many we have seen at this symposium and is not presently
the site of active SF. The models, projected into the future, indicate
that it will grow!

The feedback models yield a bit more SF in the core of the primary
than observed, but otherwise with the same widely distributed SF in
the primary, and concentration of SF in the ocular and tidal tail of
the companion. The models suggest some interesting patterns in the SF
in the primary over the course of the interaction, but these must be
further analyzed.

\section{Conclusions}

What have we learned from these and other published case studies, and
how do they complement the statistical studies? In the area of mass
transfer, the generally good agreement between model and observational
gas distributions in two very different cases - Arp 284 (extreme gas
removal from the disks) and NGC 2207 (little perturbation of the
primary disk) - is very encouraging. These general results also agree
with expectations derived from exploratory model grids. The more
detailed comparisons to observation in these two systems give us
confidence that hydrodynamic models can quite accurately reproduce
details of collisional morphology and kinematics on scales of about a
few kpc. This could motivate further checks of model predictions,
e.g., a metallicity study of the countertails of NGC 7714 to see if
there is evidence of differences that might be expected if the inner
tail gas came from the companion. Such specific predictions are not
possible without detailed modeling of individual systems.

In terms of SF, case studies are required to model modes of SF that
are either unique to a specific system, or nearly so. Possible
examples from the cases above include the inner SW tail of NGC 7714
and, the line of SF regions on the northern edge of the Arp 284
bridge. Less unique examples, but still with system specific
characteristics include the absence of SF in the NGC 7714 ring, SF in
the rim of the IC 2163 ocular, and the scattered SF in the NGC 2207
disk.

The bulk of interaction induced SF in the universe occurs in merger
remnants and the cores of unmerged collision partners. Statistical
studies are ideal for studying the mean characteristics of this type
of SF, and case studies would contribute little if this SF has a large
stochastic component. However, understanding anomalous SF, like the
examples of the previous paragraph, may be essential to understanding
the formation of globular clusters and tidal dwarfs. These are
minority populations, but still very interesting.  In fact, answering
the questions of how tidal dwarfs and globulars form, and also
understanding the sytematics of wave induced SF in disks will
require many detailed case studies.

\end{document}